\documentclass{Interspeech}

\interspeechcameraready

\title{Multichannel Keyword Spotting for Noisy Conditions}

\author[affiliation={1}]{Dzmitry}{Saladukha}
\author[affiliation={2}]{Ivan}{Koriabkin}
\author[affiliation={1}]{Kanstantsin}{Artsiom}
\author[affiliation={3}]{Aliaksei}{Rak}
\author[affiliation={3}]{Nikita}{Ryzhikov}

\affiliation{Embedded Voice Input Team}{Yandex}{Belarus}
\affiliation{Embedded Voice Input Team}{Yandex}{Serbia}
\affiliation{Embedded Voice Input Team}{Yandex}{Russia}
\email{dsoloduha@yandex-team.ru, koryabkin@yandex-team.ru, konstartyom@yandex-team.ru, alexrak@yandex-team.ru, nryzhikov@yandex-team.ru}
\keywords{speech recognition, human-computer interaction, computational paralinguistics}

\usepackage{comment}

\begin{document}

\maketitle

\begin{abstract}
    
This article presents a method for improving a keyword spotter (KWS) algorithm in noisy environments. Although beamforming (BF) and adaptive noise cancellation (ANC) techniques are robust in some conditions, they may degrade the performance of the activation system by distorting or suppressing useful signals. The authors propose a neural network architecture that uses several input channels and an attention mechanism that allows the network to determine the most useful channel or their combination. The improved quality of the algorithm was demonstrated on two datasets: from a laboratory with controlled conditions and from smart speakers in natural conditions. The proposed algorithm was compared against several baselines in terms of the quality of noise reduction metrics, KWS metrics, and computing resources in comparison with existing solutions.
\end{abstract}

\section{Introduction}

Smart speakers have become increasingly popular in recent years due to the improvement in the quality of their work and the development of their capabilities. Voice activation technology is a vital component of such devices; consequently, it has also improved recently. Currently, various approaches have been developed to tackle the problem of voice activation. The articles \cite{svdf}, \cite{cnnspotter}, \cite{hmm} demonstrate a sufficient level of voice activation quality in clean environments without significant sound distortions.

However, the KWS task under noisy conditions remains a challenge. The authors of the articles \cite{googleasrsnri}, \cite{nastar}, \cite{nrandasr} propose to use training data augmentation techniques. It mitigates some aspects of these challenges, but does not fully resolve issues related to noisy conditions. The authors of \cite{asr_denoiser} propose to use single-channel denoising, that requires complex training in conjunction with the KWS model. In the case of on-device voice activation, it is possible to use several microphone channels, which avoids the problem of network bandwidth limitations. Similar to the way humans use a binaural mechanism to better focus on specific sounds, a multichannel algorithm could partially ignore the noise component.

Currently, there is no standardized method for utilizing multiple microphones for the KWS task. The authors of \cite{attention} propose a solution that utilizes the attention mechanism to aggregate raw signal from several microphones. The authors of \cite{ji2020integration} also suggest using the attention mechanism in various directions of the minimum variance distortionless response (MVDR) beamformer in conjunction with the omni-microphone. In \cite{beamlstm}, the directions of the MVDR beamformer are determined using the long short-term memory model. The authors of \cite{hotword3mic} propose using their ANC approach and two KWS models to process both uncleaned and denoised microphone signals. The use of three-dimensional (3D) singular value decomposition filter (SVDF) and 3D convolutional neural network (CNN) architectures for processing multichannel microphone data was proposed in \cite{svdf3d} and \cite{ganapathy20183}, respectively. In \cite{multichannel-voice-trigger} the authors proposed utilizing source separation channels and aggregate them using a Transform-average-concatenate (TAC) layer.

In this paper, we present a novel approach that combines multichannel noise reduction algorithm from \cite{hotword3mic} with an attention mechanism from \cite{attention} in the KWS architecture, allowing the model to dynamically switch between input channels. To evaluate this model against several baselines, we collected the datasets in both controlled laboratory environments and real-world smart speaker operations. Our approach outperformed all baselines in terms of reliability, demonstrating superior performance in both experimental and real-world conditions. Additionally, the attention approach proved to be significantly more computationally efficient than the ensembling method.

\section{Noise reduction}
    KWS on smart devices is a relatively new area of application for noise reduction algorithms with its own peculiarities, and a unified approach has not yet been formed. In this article, we compare two methods: static beamformers with predefined directivities and ANC, which were used to improve the quality of KWS algorithms.
    
    \subsection{Microphone array}
        A specific smart speaker model was used to evaluate and test algorithms. The device features a microphone array of 7 microphones, one in the center, and six arranged in a circle with a radius of 4.2 cm with a step of 60 degrees. This array can produce beams of complex directivity, with the central microphone serving as the omnidirectional channel.

    \subsection{Directed beamformers approach}  
        Ji et al. \cite{ji2020integration} proposed a multichannel KWS model that uses five channels: four beams that cover a 360-degree segment around the smart speaker for noise reduction and an omni channel as a non-distorted signal for efficient operation in high signal-to-noise ratio (SNR) cases. In this article, we adopted the same idea as one of the baseline algorithms.

        Since we experimented with a microphone array of seven microphones, we used six beams instead of four to cover 360 degrees around the device.

    \subsection{ANC}  
        Huang et al. \cite{hotword2mic} proposed a method to enhance the robustness of KWS in noisy environments by utilizing two microphone channels. They modified the ANC technique by classifying directivity of all sounds recorded more than one second before the activation moment as directivity of noise to be suppressed. Later, they improved this algorithm in \cite{hotword3mic} to extend its applicability to multiple microphone channels, and found that the 3-microphone scheme is the most effective. In this paper, we adopt the 3-microphone algorithm as ANC approach to be evaluated. However, it is worth noting that this method may inadvertently suppress the activation word signal in case of high SNR or if the user speaks anything prior to the activation.

\section{Network Architecture}
    \subsection{Base network}
        As a baseline, we use a one-channel KWS model that works on an omni channel with an SVDF-like architecture \cite{svdf}. This model works on a sequence of sound frames with a size of 25 milliseconds and a shift of 10 milliseconds. The features pipeline calculates 40 log Mel-filterbank features for each frame. This model contains approximately 750,000 parameters and predicts the probability of the keyword for each frame.

    \subsection{Ensemble}
\begin{figure}
        \centering
        \includegraphics[scale=0.18]{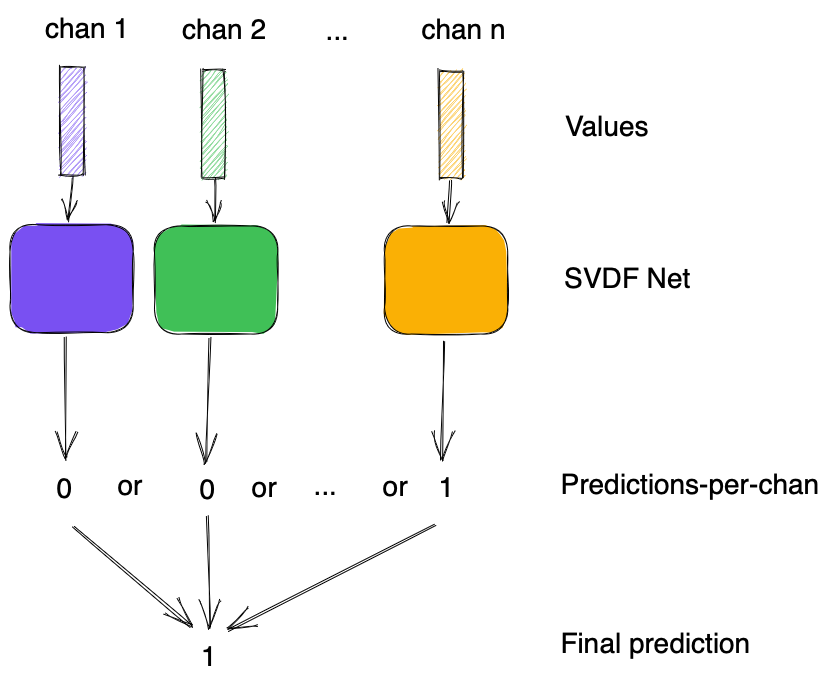}
        \caption{Scheme of ensemble with multiple spotter models for multichannel input}
        \label{fig:ensemble}
        \end{figure}

        For the ensemble, we trained a base model using omni channel data, then fine-tuned this model for a new channel.
        So, for the omni+ANC model, we fine-tuned the base model for the ANC channel. At inference time, we run the base and fine-tuned models on two distinct channels and aggregate predictions using logical OR. An activation is produced when at least one of the models predicts a keyword with a probability greater than a certain threshold for a particular channel. The block schematic of the ensemble model is shown in Figure~\ref{fig:ensemble}.

    \subsection{Attention}
    \label{attention-arch}
\begin{figure}
        \centering
        \includegraphics[scale=0.17]{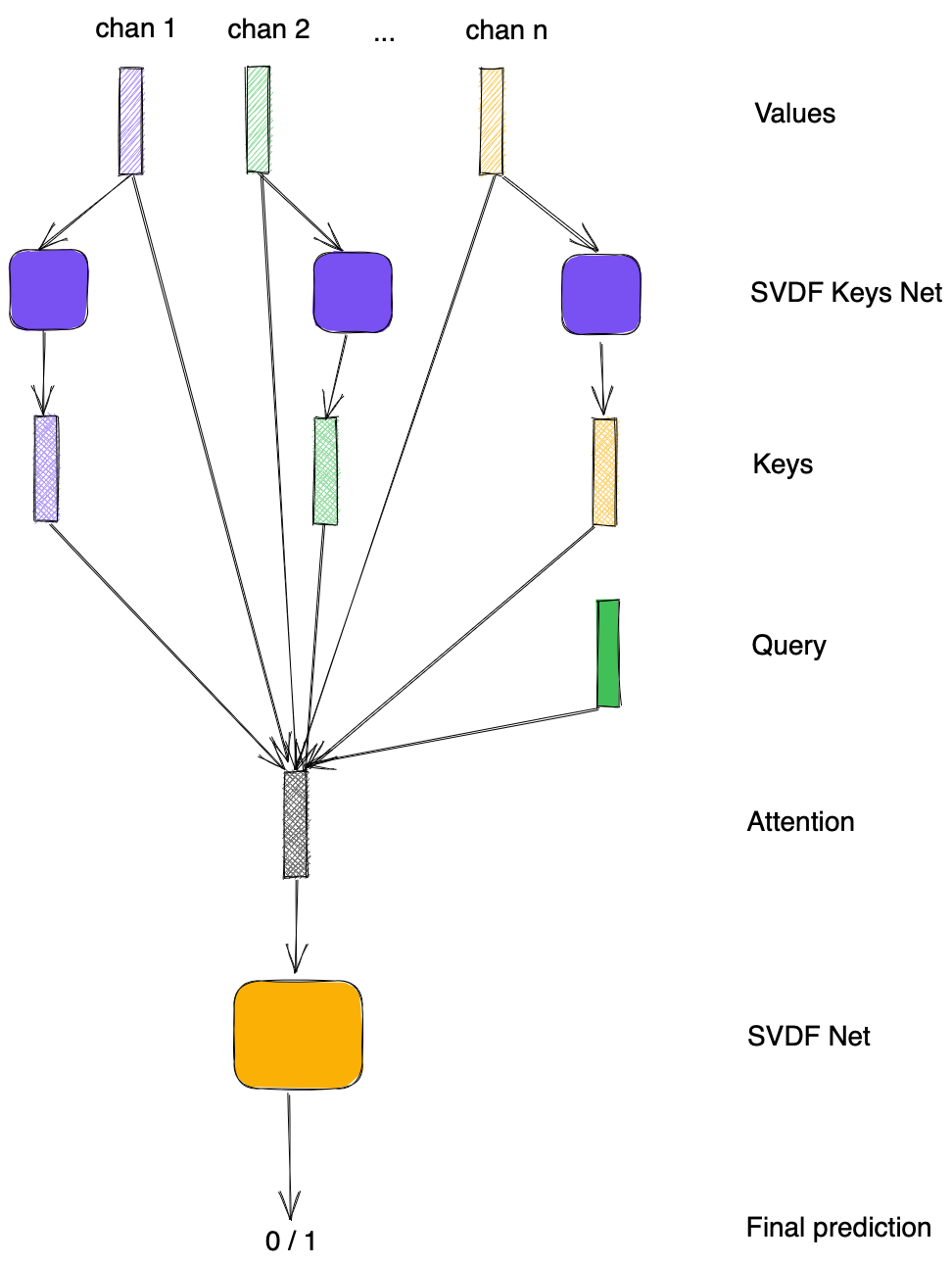}
        \caption{The architecture of spotter model extended with attention network for multichannel input}
        \label{fig:attention}
        \end{figure}

        To apply the attention mechanism to multiple channels, we use the implementation from \cite{ji2020integration}. It works by adding a small attention network $\psi(z)$ just before the base network. For each input feature $z_{ij}$, where $i$ corresponds to the channel index and $j$ to the feature index, the attention network predicts the logit $e_{ij} = \psi_j(z_i)$. The final channel $z^{*}$ that is fed into the base network is obtained by weighing the features along the channel dimension:
        \begin{gather}
            \alpha_{ij} = \frac{\exp{e_{ij}}}{\sum_{i} \exp{e_{ij}}} \\
            z^{*}_{j} = \sum_{i} \alpha_{ij} z_{ij}
        \end{gather}
        The attention network has an architecture similar to the base network, but with a much smaller parameter count. In our case, the attention network contains approximately 50,000 parameters (SVDF Keys Net on Figure~\ref{fig:attention}).

\section{Datasets}
        To evaluate the quality of ANC algorithms, two private datasets were utilized, each recorded in a different scenario. The first dataset was recorded in an acoustic laboratory with low levels of reverberation and residual noise, and is primarily used to accurately measure acceptable SNR levels in a controlled environment. The second dataset was created from end-user recordings and was used for training and evaluating KWS models. More detailed information regarding the dataset configurations is described in the following sections.
        
    \subsection{Acoustic Laboratory dataset}
        The acoustic laboratory used for dataset recording is a $4.5\text{m} \times 5.5\text{m}$ sound-isolated room designed to have a flat impulse response. The device being tested is placed in the center of the room on a flat platform 90cm above the floor level. Several sound sources are placed in a circle with a 2m radius around the device at different heights (70-120cm from the ground) and angles. Each recording has one sound source playing a noise that has been normalized to have a volume of 60dB SPL(A) on an external microphone in the center of the room, and one sound source at the same time is playing a keyword sound with volumes uniformly distributed from [35, 40, 45, 50, 55, 60, 65] dB SPL(A). Each recording has only one of the pre-recorded keyword sounds and uses only one of the pre-recorded types of noise: kitchen, street, vacuum cleaner, white, pink, or TV. Keyword sounds use male, female, and children's voices in equal proportions. The resulting dataset has 900 records with SNR varying from [-25, -20, -15, -10, -5, 0, 5] dB SPL(A).

    \subsection{KWS dataset}
    \label{subsubsection:dataset}
        We used large in-house multichannel partially human-annotated and partially pseudo-labeled \cite{pseudolabels} dataset to train and evaluate KWS model.

        The training data for the experiments described below consists of 500,000 pseudo-labeled multichannel utterances. The testing data for results reported below consists of 50 thousands human-transcribed multichannel utterances and contains 40 thousands target keywords.

        Also we have 10 millions omni-channel pseudo-labeled utterances.
        
\section{Evaluation}
    \subsection{Evaluated approaches}
        We evaluated six multichannel approaches for KWS:
        \begin{itemize}
            \item \textit{Base KWS} \cite{svdf}: Single-channel baseline model and omni channel;
            \item \textit{Base KWS + ANC}: Single-channel baseline model and ANC channel;
            \item \textit{Base KWS x2}: Single-channel baseline model doubled in size and omni channel
            \item \textit{Ensemble KWS + ANC} \cite{hotword3mic}: Ensemble of base KWS models with ANC and omni channel;
            \item \textit{Attention KWS + BF} \cite{ji2020integration}: KWS model with attention mechanism with 6 channels from BF algorithm and omni channels;
            \item \textbf{\textit{Attention KWS + ANC (proposed)}}: KWS model with attention mechanism with ANC and omni channels.
        \end{itemize}

        The ensemble method for the BF algorithm was not studied since it is computationally complex to run the base model on each BF channel.
        
    \subsection{Metrics}
        \subsubsection{KWS metrics}
        We use false reject rate (FRR) and false alarms per hour (FA/h) metrics to evaluate KWS models. These metrics are not independent, so improving one metric by changing the threshold can lead to a decline in the other. To compare different models, we adjust them to FA/h = 0.1 and compare them using the FRR metric.
        
        \subsubsection{Noise cancellation evaluation}
\begin{figure}[b]
        \includegraphics[scale=0.4]{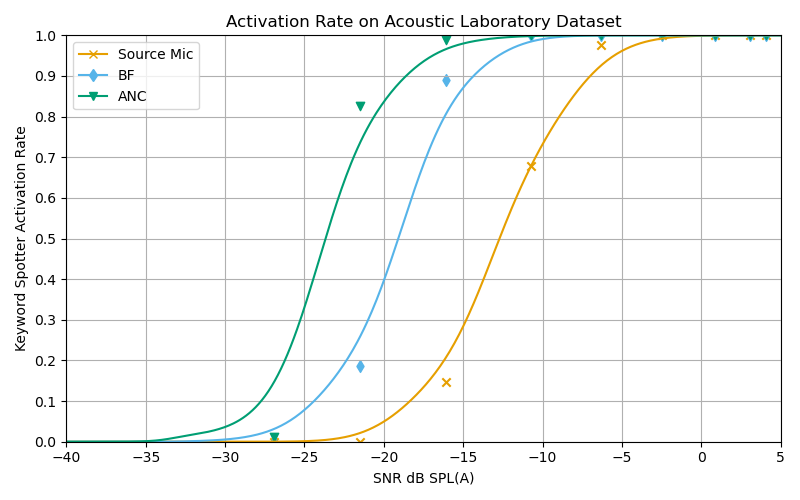}
        \caption{SNR versus spotter activation rate plot for laboratory data}
        \label{fig:snr_lab}
        \end{figure}

        To evaluate the efficiency of the noise cancellation algorithm on the laboratory dataset, the base KWS model was applied to the BF, ANC and omni channels. Using the estimated values of SNR, we could evaluate how the KWS model's FRR depends on the SNR level, as shown in Figure~\ref{fig:snr_lab}. To achieve numerical values, we interpolated curves, and the SNR gain between omni channel and the measured algorithm at a specific level of FRR was used. This way, we can estimate how much SNR improvement could be achieved for this algorithm. In the case of BF, we used the oracle approach, i.e., the best KWS confidence of multiple channels was taken for evaluation. The results of the measurements are presented in Table~\ref{table:snr_lab_table}.

    %    \begin{table}[th]
    %    \caption{SNR levels for voice activation with 50\%, 30\% and 10\% FRR on laboratory data}
    %    \centering
    %    \begin{tabular}{cccc}
    %    \toprule
    %    & \textbf{FRR=50\%} & \textbf{FRR=30\%} & \textbf{FRR=10\%} \\ 
    %    \midrule
    %    omni & -12.7 & -10.4 & -7.0 \\
    %    BF & -19.1 & -17.3 & -14.3 \\
    %    \textbf{ANC} & \textbf{-23.7} & \textbf{-21.9} & \textbf{-18.6} \\
    %    \bottomrule
    %    \end{tabular}
    %    \label{table:snr_lab_table}
    %    \end{table}

        \begin{table}[th]
        \caption{FRR (\%) levels on different SNR conditions recorded in the acoustic laboratory}
        \centering
        \begin{tabular}{ccccc}
        \toprule
        \textbf{SNR} & \textbf{-10dB} & \textbf{-15dB} & \textbf{-20dB} & \textbf{-25dB} \\ 
        \midrule
        Source Mic & 26.7 & 71.2 & 95.3 & 100.0 \\
        BF & 0.8 & 13.5 & 59.8 & 92.9 \\
        \textbf{ANC} & \textbf{0.1} & \textbf{2.8} & \textbf{16.8} & \textbf{67.1} \\
        \bottomrule
        \end{tabular}
        \label{table:snr_lab_table}
        \end{table}

\section{Experiments}

    \subsection{Training}
        \subsubsection{Baseline training}
        We trained our model using setup similar to encoder+decoder from \cite{svdf} on omni channel data. Also we add max-pool per utterance loss to improve quality of the model.

        Max-pool loss by utterances for one-phrase KWS model:
        \begin{gather}
            \hat{p} = \prod_{p \in parts_k} \max_{t} x_{tp} \\
            \mathcal{L}_{mp} = - (1 - y) \log{(1 - \hat{p})} - y \log{\hat{p}}
        \end{gather}

        \noindent
        where $T$ is the number of frames in the utterance, $x_{ij}$ is the probability of the $j$th class on the $i$th frame, $y_{i}$ is the target class on the $i$th frame, $y$ is the whole utterance target, $parts_k$ are the phonemes that make up the target phrase  for encoder or the whole phrase for decoder.

        \subsubsection{Baseline KWS model fine-tuning}
        As our KWS dataset predominantly comprises omni channel data (Section~\ref{subsubsection:dataset}), the initial model is trained on such data and than being fine-tuned for new channel data. 

        \subsubsection{Ensemble KWS model fine-tuning}
        Initially, the ensemble model produced more false alarms than the baseline model because of the logical OR of outputs from per-channel models. To tune this model to fixed FA/h and best FRR, we use grid-search by per-channel thresholds to find the best values. Often after grid-search we have best result with different thresholds per channel because audio data can be strongly different.
        
        \subsubsection{Attention KWS model fine-tuning}
        To achieve the architecture described in \ref{attention-arch}, we began with training the model with omni channel data, then added the attention network to the input of the base network and fine-tuned model with setup similar to baseline training on multichannel data.

    \subsection{Hardware benchmarks}
        To run benchmarks, we use a smart speaker with Amlogic S905X2 socket with 4 core ARMv8 Cortex-A53 1.8 GHz, and for reference, a regular PC with Intel Core i7-3540M. Noise reduction algorithm and keyword spotting each use a single CPU core.
        
        \begin{table}[th]
        \caption{Noise reduction resource usage: real-time factor metric}
        \centering
        \begin{tabular}{ccc}
        \toprule
        & \textbf{S905X2} & \textbf{i7-3540M} \\ \midrule
        BF & 0.112 & 0.015 \\ 
        \textbf{ANC} & \textbf{0.070} & \textbf{0.010} \\ \bottomrule
        \end{tabular}
        \label{table:vqe_perf}
        \end{table}
        \begin{table}[th]
        \caption{Keyword spotter resource usage}
        \centering
        \setlength{\tabcolsep}{4pt} % Reduce column padding
        \begin{tabular}{cccc}
        \toprule
        & \multicolumn{2}{c}{\textbf{Real-time factor}} & \textbf{Model size} \\
        & \textbf{S905X2} & \textbf{i7-3540M} & \textbf{(MB)} \\ \midrule
        Base KWS & \textbf{0.18} & \textbf{0.020} & \textbf{3.4} \\
        Base KWS x2 & 0.34 & 0.034 & 6.9 \\
        Ensemble KWS + ANC & 0.36 & 0.041 & 6.6 \\
        Attention KWS + ANC & 0.21 & 0.026 & 3.5 \\
        Attention KWS + BF & 0.30 & 0.043 & 3.5 \\ \bottomrule
        \end{tabular}
        \label{table:spotter_perf}
        \end{table}

        According to measurements in Table~\ref{table:vqe_perf}, all voice enhancement algorithms are viable to use on an ARM CPU. The difference in the RSS memory usage of those variants is negligible, with all using approximately 10MB. Table~\ref{table:spotter_perf} shows that all KWS models can be used on consumer devices, although both the real-time factor and the model size of ensemble and x2 models are significantly higher. On the other hand, the proposed solution introduces only a small overhead compared to the baseline.

\section{Results and Discussion}

    \begin{table}[th]
    \caption{False Reject Rate at fixed false alarms per hour level on a test dataset}
    \centering
        \begin{tabular}{cc}
             \toprule
             \textbf{Model} & \textbf{FRR (\%)}  \\ \midrule
             Base KWS & 7.5 \\
             Base KWS + ANC & 8.8 \\
             Base KWS x2 & 7.0 \\
             Ensemble KWS + ANC & 6.4 \\ 
             Attention KWS + BF & 6.7 \\ 
             \textbf{Attention KWS + ANC} & \textbf{5.5} \\ \bottomrule
        \end{tabular}
    \label{table:frr_table}
    \end{table}

    Table~\ref{table:frr_table} shows that keyword spotting only with the ANC channel alone did not perform well on the test set, likely due to the mentioned downside of the algorithm, when the signal is misinterpreted as noise. Using noise reduction algorithms like ANC or beamforming in combination with an omni channel, on the other hand, significantly improves FRR, outperforming even doubling the model size. These results suggest that incorporating additional channels with noise reduction processing carries valuable information under low SNR conditions while allowing the model to use nondistorted audio if SNR is high.
    
    Our proposed Attention KWS + ANC approach achieved the lowest FRR, significantly outperforming both the beamforming-based model and the ensemble model with ANC. Most probably, beamforming was surpassed due to better noise reduction on the ANC channel, as shown in Table~\ref{table:snr_lab_table}. The advantage of attention-based processing over ensembling may lie in its ability to leverage not only individual channels but also their weighted combinations, improving the model’s accuracy.

    At the same time, the proposed algorithm is CPU- and memory-efficient, making it the most favorable solution for real-world KWS applications.

    \section{Conclusion and Future Work}
    We propose a multichannel KWS approach that leverages the advantages of the ANC channel to improve KWS models. Our training setup enables the derivation of a multichannel model from a single-channel baseline. Experimental results demonstrated that the proposed attention model significantly outperforms other methods, achieving the lowest false reject rate while maintaining computational efficiency suitable for on-device deployment.
    
    For further research, it may be interesting to select the most effective channel or the combination of channels to transmit from the device to the cloud, in order to enhance the quality of automatic speech recognition (ASR) models and other components of the voice assistant. This would allow the ASR models to benefit from multichannel noise reduction without increasing the bandwidth load. While this task can be easily accomplished with the ensemble model, it poses a greater challenge with the attention model.

\bibliographystyle{IEEEtran}
\bibliography{mybib}

\begin{thebibliography}{10}
\providecommand{\url}[1]{#1}
\csname url@samestyle\endcsname
\providecommand{\newblock}{\relax}
\providecommand{\bibinfo}[2]{#2}
\providecommand{\BIBentrySTDinterwordspacing}{\spaceskip=0pt\relax}
\providecommand{\BIBentryALTinterwordstretchfactor}{4}
\providecommand{\BIBentryALTinterwordspacing}{\spaceskip=\fontdimen2\font plus
\BIBentryALTinterwordstretchfactor\fontdimen3\font minus \fontdimen4\font\relax}
\providecommand{\BIBforeignlanguage}[2]{{%
\expandafter\ifx\csname l@#1\endcsname\relax
\typeout{** WARNING: IEEEtran.bst: No hyphenation pattern has been}%
\typeout{** loaded for the language `#1'. Using the pattern for}%
\typeout{** the default language instead.}%
\else
\language=\csname l@#1\endcsname
\fi
#2}}
\providecommand{\BIBdecl}{\relax}
\BIBdecl

\bibitem{svdf}
R.~Alvarez and H.-J. Park, ``End-to-end streaming keyword spotting,'' in \emph{ICASSP 2019 - 2019 IEEE International Conference on Acoustics, Speech and Signal Processing (ICASSP)}, 2019, pp. 6336--6340.

\bibitem{cnnspotter}
T.~Sainath and C.~Parada, ``Convolutional neural networks for small-footprint keyword spotting,'' in \emph{Interspeech}, 2015.

\bibitem{hmm}
R.~Rose and D.~Paul, ``A hidden markov model based keyword recognition system,'' in \emph{International Conference on Acoustics, Speech, and Signal Processing}, 1990, pp. 129--132 vol.1.

\bibitem{googleasrsnri}
\BIBentryALTinterwordspacing
Y.~Koizumi, S.~Karita, A.~Narayanan, S.~Panchapagesan, and M.~Bacchiani, ``Snri target training for joint speech enhancement and recognition,'' 2022. [Online]. Available: \url{https://arxiv.org/abs/2111.00764}
\BIBentrySTDinterwordspacing

\bibitem{nastar}
\BIBentryALTinterwordspacing
C.-C. Lee, C.-H. Hu, Y.-C. Lin, C.-S. Chen, H.-M. Wang, and Y.~Tsao, ``Nastar: Noise adaptive speech enhancement with target-conditional resampling,'' 2022. [Online]. Available: \url{https://arxiv.org/abs/2206.09058}
\BIBentrySTDinterwordspacing

\bibitem{nrandasr}
N.~Howard, A.~Park, T.~Z. Shabestary, A.~Gruenstein, and R.~Prabhavalkar, ``A neural acoustic echo canceller optimized using an automatic speech recognizer and large scale synthetic data,'' in \emph{ICASSP 2021 - 2021 IEEE International Conference on Acoustics, Speech and Signal Processing (ICASSP)}, 2021, pp. 7128--7132.

\bibitem{asr_denoiser}
\BIBentryALTinterwordspacing
S.~E. Eskimez, X.~Wang, M.~Tang, H.~Yang, Z.~Zhu, Z.~Chen, H.~Wang, and T.~Yoshioka, ``Human listening and live captioning: Multi-task training for speech enhancement,'' 2021. [Online]. Available: \url{https://arxiv.org/abs/2106.02896}
\BIBentrySTDinterwordspacing

\bibitem{attention}
\BIBentryALTinterwordspacing
H.~Zhang, J.~Zhang, and Y.~Wang, ``End-to-end models with auditory attention in multi-channel keyword spotting,'' 2018. [Online]. Available: \url{https://arxiv.org/abs/1811.00350}
\BIBentrySTDinterwordspacing

\bibitem{ji2020integration}
X.~Ji, M.~Yu, J.~Chen, J.~Zheng, D.~Su, and D.~Yu, ``Integration of multi-look beamformers for multi-channel keyword spotting,'' in \emph{ICASSP 2020 - 2020 IEEE International Conference on Acoustics, Speech and Signal Processing (ICASSP)}, 2020, pp. 7464--7468.

\bibitem{beamlstm}
T.~N. Sainath, R.~J. Weiss, K.~W. Wilson, B.~Li, A.~Narayanan, E.~Variani, M.~Bacchiani, I.~Shafran, A.~Senior, K.~Chin, A.~Misra, and C.~Kim, ``Multichannel signal processing with deep neural networks for automatic speech recognition,'' \emph{IEEE/ACM Transactions on Audio, Speech, and Language Processing}, vol.~25, no.~5, pp. 965--979, 2017.

\bibitem{hotword3mic}
Y.~Huang, T.~Z. Shabestary, A.~Gruenstein, and L.~Wan, ``Multi-microphone adaptive noise cancellation for robust hotword detection,'' in \emph{Proc. InterSpeech 2019}, 2019, pp. 1233--1237.

\bibitem{svdf3d}
J.~Wu, Y.~Huang, H.-J. Park, N.~Subrahmanya, and P.~Violette, ``Small footprint multi-channel keyword spotting.'' in \emph{Odyssey}, 2020, pp. 391--395.

\bibitem{ganapathy20183}
S.~Ganapathy and V.~Peddinti, ``3-d cnn models for far-field multi-channel speech recognition,'' in \emph{2018 IEEE International Conference on Acoustics, Speech and Signal Processing (ICASSP)}, 2018, pp. 5499--5503.

\bibitem{multichannel-voice-trigger}
\BIBentryALTinterwordspacing
T.~Higuchi, A.~Brueggeman, M.~Delfarah, and S.~Shum, ``Multichannel voice trigger detection based on transform-average-concatenate,'' 2024. [Online]. Available: \url{https://arxiv.org/abs/2309.16036}
\BIBentrySTDinterwordspacing

\bibitem{hotword2mic}
Y.~A. Huang, T.~Z. Shabestary, and A.~Gruenstein, ``Hotword cleaner: Dual-microphone adaptive noise cancellation with deferred filter coefficients for robust keyword spotting,'' in \emph{ICASSP 2019 - 2019 IEEE International Conference on Acoustics, Speech and Signal Processing (ICASSP)}, 2019, pp. 6346--6350.

\bibitem{pseudolabels}
\BIBentryALTinterwordspacing
H.-J. Park, P.~Zhu, I.~L. Moreno, and N.~Subrahmanya, ``Noisy student-teacher training for robust keyword spotting,'' 2021. [Online]. Available: \url{https://arxiv.org/abs/2106.01604}
\BIBentrySTDinterwordspacing

\end{thebibliography}

\end{document}